\documentclass[pre,aps,superscriptaddress,showpacs]{revtex4-1}
\usepackage{natbib}
\usepackage{amsmath}
\usepackage{epsfig}
\usepackage{color}

\DeclareMathOperator*{\sumint}{%
\mathchoice%
{\ooalign{$\displaystyle\sum$\cr\hidewidth$\displaystyle\int$\hidewidth\cr}}
{\ooalign{\raisebox{.14\height}{\scalebox{.7}{$\textstyle\sum$}}\cr\hidewidth$\textstyle\int$\hidewidth\cr}}
{\ooalign{\raisebox{.2\height}{\scalebox{.6}{$\scriptstyle\sum $}}\cr$\scriptstyle\int$\cr}}
{\ooalign{\raisebox{.2\height}{\scalebox{.6}{$\scriptstyle\sum$}}\cr$\scriptstyle\int$\cr}}
}

\begin{document}

\title{On a tilted Liouville-master equation of open quantum systems}
\author{Fei Liu}
\email[Email address: ]{feiliu@buaa.edu.cn}
\affiliation{School of Physics, Beihang University, Beijing 100191, China}

\date{\today}

\begin{abstract}
{A tilted Liouville-master equation in Hilbert space is presented for Markovian open quantum systems. We demonstrate that it is the unraveling of the tilted quantum master equation. The latter is widely used in the analysis and calculations of stochastic thermodynamic quantities in quantum stochastic thermodynamics.  }
\end{abstract}
\maketitle

\section{Introduction}
\label{section1}
In the past two decades, stochastic thermodynamics for open quantum systems has attracted considerable theoretical interest~\cite{Esposito2009,Campisi2011,Alicki2018,Liu2018}. One of the major issues is the statistics of random thermodynamic variables such as heat, work, entropy production, and efficiency~\cite{Kurchan2000,Breuer2003,Talkner2007,DeRoeck2007,Talkner2008,Crooks2008,
Garrahan2010,Subasi2012,Horowitz2012,Hekking2013,Leggio2013, Horowitz2013,Zinidarifmmodeheckclseci2014, Verley2014a,Gasparinetti2014,Cuetara2015,Carrega2015,Manzano2016,Suomela2016,Liu2016a, Strasberg2017,Wang2017,Restrepo2018,Carollo2018,Carollo2019,Liu2020}. The tilted or generalized quantum master equation (TQME) is a useful approach for the study of these problems~\cite{Esposito2009}. For instance, the fluctuation theorems of steady-states can be demonstrated according to the symmetries implied in the maximal eigenvalue of the equation~\cite{Esposito2009,Gasparinetti2014,Cuetara2015,Liu2020}. To study the concrete probability distributions of the random thermodynamic variables, we can numerically or analytically solve the equation to obtain characteristic functions or moment generating functions~\cite{Silaev2014,Gasparinetti2014,Cuetara2015,Carrega2015,Liu2016a,Wang2017,Restrepo2018}; the distributions and functions are mathematically equivalent.

TQME was developed by Esposito et al.~\cite{Esposito2009} in their investigation of the fluctuation theorems of open quantum systems. Their derivation is based on the celebrated two-energy measurement scheme~\cite{Kurchan2000} and the whole procedure is similar to the derivation of the quantum master equation~\cite{Alicki2006,Rivas2012,Breuer2000}. Indeed, these two equations appear to be almost the same in form; the only difference is an exponential term in the front of the jump terms therein. Although TQME is established in stochastic thermodynamics, an analogous equation was presented by Mollow in 1975 in a seminal paper about photon emissions of quantum systems with discrete energy levels~\cite{Mollow1975}. Importantly, the notion of quantum jumps was proposed in this paper~\cite{Carmichael1989,Gardiner2004}. To determine the photon-number distributions in the many modes of fluorescence, Mollow developed a hierarchy of equations. Garrahan and Lesanovsky argued that these equations are equivalent to TQME~\cite{Garrahan2010}. This result inspired us to re-derive TQME from the perspective of quantum jumps~\cite{Liu2016a,Liu2018}. Different from the method of Esposito et al.~\cite{Esposito2009}, our method is based on the unraveling of the quantum master equation into quantum jump trajectories~~\cite{Breuer2000}, and Dyson expansion is used~\cite{Kist1999,Liu2016a}.

\begin{figure}
\includegraphics[width=1.\columnwidth]{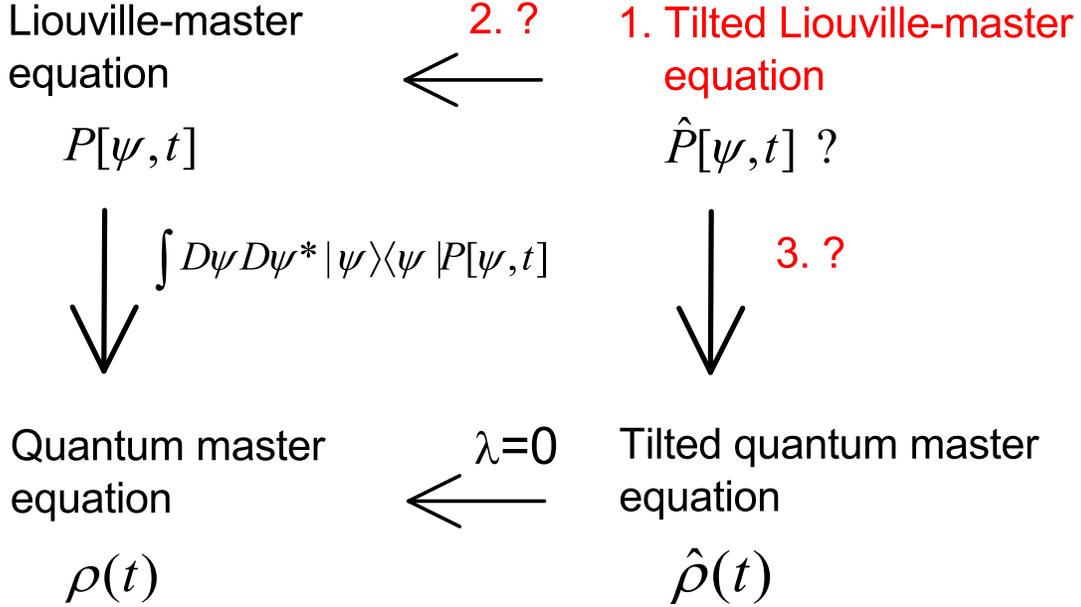}
\caption{Three questions of interest addressed in this paper: 1. Is there a tilted Liouville-master equation in the Hilbert space? 2. What is the relationship between it and the Liouville-master equation? 3. Does the equation yield the conventional TQME?  }
\label{fig1}
\end{figure}

Although TQME is widely adopted in the literature, several open questions still remain. Quantum jump trajectories, or trajectories for short, are composed of alternating deterministic pieces and random jumps of wave functions of individual quantum systems. They are a type of stochastic process called the piecewise deterministic process (PDP)~\cite{Davis1984}. Hence, in Hilbert space, there exists a Liouville-master equation that governs the dynamics of the probability distribution of the random wave functions. Moreover, this equation exactly yields the quantum master equation for the reduced density matrix ~\cite{Breuer1995a,Breuer1995,Breuer1997,Breuer1997a}. This naturally raises the following questions: Is there a tilted Liouville-master equation in the Hilbert space?  What is the relationship between the equation and the Liouville-master equation? Does it yield TQME? We also list these questions in Fig.~(\ref{fig1}). The aim of this paper is to present definite answers to these questions.

The remainder of this paper is organized as follows: In Sec.~(\ref{section2}), the Markovian quantum master equation and its unraveling of trajectories are briefly reviewed. In Sec.~(\ref{section3}), by defining two functionals, we rewrite the probability distribution functional of the trajectory in a new form. In Sec.~(\ref{section4}), a Feynman-Kac formula in the Hilbert space is derived. Based on these two results, in Sec.~(\ref{section5}), we derive the tilted Liouville-master equation and demonstrate that it is indeed the unraveling of TQME. Section~(\ref{section6}) concludes the paper.

\section{Unraveling of Markovian quantum master equation}
\label{section2}
We start with the Markovian quantum master equation. Because our aim is to prepare the essential equations and notations, the description is brief; a detailed explanation can be found in the standard textbooks, e.g., Ref.~\cite{Breuer2000}. Let $\rho(t)$ be the reduced density matrix of an open quantum system. Under appropriate assumptions and conditions, the ensemble dynamics of the system is described by the Markovian quantum master equation~\cite{Davies1974,Lindblad1976,Gorini1976}
\begin{eqnarray}
\label{quantummasterequation}
\partial_t \rho(t)=-i[H,\rho(t)]+\sum_{\alpha=1}^M r_\alpha\left( A_\alpha\rho(t)A^\dag_\alpha -\frac{1}{2}\left\{A^\dag_\alpha A_\alpha,\rho(t)\right\}\right),
\end{eqnarray}
where the Planck constant $\hbar$ is set to 1, $H$ denotes the Hamiltonian of the quantum system, $A_\alpha$ is the Lindblad operator, and nonnegative $r_\alpha$, $\alpha=1,\cdots,M$ represent the correlation functions of the environments surrounding the system. Remarkably, this general equation is also equal to an ensemble average of the density matrices of the individual stochastic quantum systems~\cite{Breuer1995a,Breuer1995,Breuer1997,Breuer1997a}:
\begin{eqnarray}
\label{densitymatrixzz}
\rho(t)=\int D\psi D\psi^* P[\psi,t] |\psi(t)\rangle\langle \psi(t)|.
\end{eqnarray}
Here, $D\psi D\psi^*$ is the Hilbert space volume element, and $P[\psi,t]$ is the probability distribution functional of the random wave function $\psi$ at time $t$. The latter satisfies the Liouville-master equation~\cite{Breuer2000,Breuer1997a}:
\begin{eqnarray}
\label{quantumLiouvillemasterequation}
\partial_t P[\psi,t]&=&i\int dz\left ( \frac{\delta}{\delta \psi(z)}G[\psi](z)-\frac{\delta}{\delta \psi^*(z)}G[\psi]^*(z) \right ) P[\psi,t]\nonumber +\nonumber \\
&&\int D\phi D\phi^*\left ( P[\phi,t]W[\phi|\psi] -P[\psi,t]W[\psi|\phi] \right ),
\end{eqnarray}
where $\delta/\delta \psi(z)$ and $\delta/\delta^* \psi(z)$ are functional derivatives and $z$ denotes the positional coordinate. The operator $G$ in the first integral of Eq.~(\ref{quantumLiouvillemasterequation}) is
\begin{eqnarray}
\label{Goperator}
G[\psi]=\left(\hat H  + \frac{i}{2}\sum_{\alpha=1}^M r_\alpha \parallel A_\alpha\psi\parallel ^2\right ) \psi,
\end{eqnarray}
and $\hat H\equiv H-({i}/{2})\sum_{\alpha=1}^M r_\alpha A_\alpha^\dag A_\alpha $ is the non-Hermitian Hamiltonian. In the second integral of Eq.~(\ref{quantumLiouvillemasterequation}), the transition rate is
\begin{eqnarray}
\label{totalrate1}
W[\phi|\psi]=\sum_{\alpha=1}^M k_
\alpha[\phi] \delta\left[\frac{A_\alpha\phi }{\parallel A_\alpha \phi \parallel }-\psi\right],
\end{eqnarray}
where $\delta[$ $]$ denotes the Dirac functional, $k_\alpha[\phi]= r_\alpha \parallel A_\alpha \phi  \parallel ^2$ is the rate of jump of the wave function from $\phi$ to the target $\psi_\alpha={A_\alpha\phi }/{\parallel A_\alpha \phi \parallel } $, and the precise definition of the latter is given in Eq.~(\ref{targetwavefunction}) below. Eq.~(\ref{quantumLiouvillemasterequation}) is called the unraveling of the quantum master equation~(\ref{quantummasterequation}).

The stochastic process underlying Eq.~(\ref{quantumLiouvillemasterequation}) is the PDP in the Hilbert space~\cite{Breuer2000}. The evolutions or trajectories of the individual quantum systems are composed of deterministic pieces and random jumps~\cite{Breuer2000}: the former are the solutions of the nonlinear Schr$\ddot{o}$dinger equation,
\begin{eqnarray}
\label{nonlinearSchrodingerequation}
\frac{d}{d\tau}\psi(\tau)  &=&-iG[\psi(\tau)],
\end{eqnarray}
while the latter are the instantaneous jumps of wave functions given by
\begin{eqnarray}
\label{targetwavefunction}
\psi(\tau)  \rightarrow \psi_\alpha=\frac{A_\alpha \psi(\tau) }{\parallel A_\alpha \psi(\tau) \parallel},
\end{eqnarray}
$\alpha=1,\cdots,M$, and the rates of the jumps are nothing but $k_\alpha[\psi(\tau)]$. The target wave functions $\psi_\alpha $ in Eq.~(\ref{targetwavefunction}) appear to depend on the concrete wave functions $\psi(\tau)$ at the jumping times. However, in the cases of physical interest, they are uniquely specified by the Lindblad operators $A_\alpha$~\cite{Breuer2000}. In this paper, we consider only such cases.

\section{Probability distribution functional of trajectory}
\label{section3}
Let us denote the trajectories of the individual quantum systems as $\Psi_t$, where $t$ denotes the duration of the quantum nonequilibrium process. We denote the jumps along a trajectory by $\alpha_i \in \{1,\cdots,M \}$, and $i$ represents the time $t_i$ in which the $i$-th jump occurs, where $i=0,1,\cdots, N$, and $N$ is the total number of jumps. Then, the probability distribution functional of monitoring a trajectory $\Psi_t$ is simply~\cite{Breuer2000}
\begin{eqnarray}
\label{probdensityfunctionaloftrajectory}
{\mathbf P}_{0}[\Psi_t]
&=& \exp\left[-\int_{0}^{t}ds  \Gamma[\psi(s)] \right]\prod_{i=1}^N k_{\alpha_i}[\psi(t_i^-) ],
\end{eqnarray}
where we use the subscript $0$ on the left-hand side to denote the initial wave function of the quantum system being in a target wave function $\psi_{\alpha_0}$, and $\psi(t_i^-)$ is the wave function of the system immediately prior to the jumping time $t_i$. The total rate of jumps is
\begin{eqnarray}
\label{totalrate2}
\Gamma[\psi]&=&\int D\phi D\phi^* W[\psi|\phi]\nonumber \\
&=&\sum_{\alpha=1}^M k_\alpha[\psi].
\end{eqnarray}
We emphasize that in the time interval between $t_i$ and $t_{i+1}$, $\psi(s)$ follows the solution of Eq.~(\ref{nonlinearSchrodingerequation}); the initial condition is specified by the target wave function after the last jump, that is $\psi(s=t_i^+)=\psi_{\alpha_i}$ and $t_i^+$ is the time immediately after the jump. The exponential decay function in Eq.~(\ref{probdensityfunctionaloftrajectory}) implies that during the successive evolutions, the quantum system has a probability of jumping from the current state to the potential target wave functions. This is the physical consequence arising from irreversible dissipations induced by environments or from continuous measurements on the individual quantum systems~\cite{Breuer2000,Wiseman2010,Plenio1998}.

The probability distribution functional of the trajectory~(\ref{probdensityfunctionaloftrajectory}) has an alternative form:
\begin{eqnarray}
\label{probdistributiontraj2}
\mathbf{P}_0[\Psi_t]&=& \exp\left (-t \int D\phi D\phi^* \Gamma[\phi]P_{\phi}[\Psi_t]+t \sum_{\alpha=1}^M\int D\phi D\phi^* \ln k_\alpha[\phi] F_{\phi,\alpha}[\Psi_t]  \right),
\end{eqnarray}
where we define two functionals,
\begin{eqnarray}
\label{empiricaldensity}
P_{\phi}[\Psi_t]&=&\frac{1}{t} \int_{0}^{t}ds \delta[\psi(s)-\phi]  ,  \\
\label{empiricalflux}
F_{\phi,\alpha}[\Psi_t]&=&\frac{1}{t}\sum_{i=1}^N \delta[\phi-\psi(t_i^-) ]\delta_{\alpha,\alpha_i},
\end{eqnarray}
and $\delta_{\alpha,\alpha_i}$ is the Kronecker delta symbol. It is not difficult to see that Eq.~(\ref{empiricaldensity}) represents the fraction of time of the quantum system occupying the wave function $\phi$ in the duration $t$, while Eq.~(\ref{empiricalflux}) is the empirical rate of jumping from the special $\phi$ to the special target $\psi_\alpha$. At long time limits, we expect that these functionals reduce to the probability distribution of wave functions and flows of the steady-state of the quantum ensemble, respectively. Eq.~(\ref{probdistributiontraj2}) can be used to obtain the rate functional of the level 2.5 large deviations of open quantum systems~\cite{Carollo2019,Carollo2021}. We do not pursue this issue in this paper.

To close this section, we introduce notation for formally expressing the ensemble average of an arbitrary functional $O[\Psi_t]$ of trajectories:
\begin{eqnarray}
\label{trajectoryensembleexpression}
\left \langle O \right\rangle\equiv\sum_{\alpha_0,\Psi_t} P[\psi_{\alpha_0}]{\bf P}_0[\Psi_t] O[\Psi_t],
\end{eqnarray}
where $P[\psi_{\alpha_0}]$ represents the initial probability distribution functional and the summations are over all possible trajectories and initial target wave functions. Although Eq.~(\ref{trajectoryensembleexpression}) is not rigorous, it is adequate to support our below discussions.

\section{A Feynman-Kac formula}
\label{section4}
In this section, we present a Feynman-Kac formula for the PDPs in the Hilbert space. The basis of our discussion is the intuitive picture of the trajectories and simple relations among the probabilities in Eqs.~(\ref{nonlinearSchrodingerequation}) and~(\ref{targetwavefunction}). Consider a time-dependent integral
\begin{eqnarray}
\label{integralfunctional}
u(t)=\int_0^t ds V[\psi(s)].
\end{eqnarray}
We assume $u(t)$ to be a continuous real function of time $t$. Obviously, the integral is also a functional of the trajectories.

Our aim is to calculate the probability distribution $p(u,t)$ of the random variable $u(t)$. To this end, we need several intermediate quantities and equations. First, we define $P[\psi,u,t]$ as the probability distribution functional of finding the individual quantum systems with wave function $\psi$ and the integral~(\ref{integralfunctional}) equal to $u$ at time $t$. We have
$p(u,t)=\int D\psi D\psi^* P[\psi,u,t]$. Moreover, we define again another functional $P[\psi,u,t;\psi_\alpha,v,\tau]$: it represents the joint probability distribution functional when the wave function of the quantum system and the integral at time $t$ are $\psi$ and $u$, respectively, the successive time of evolution is $\tau$, the target wave function and the integral at the time of last jump are equal to $\psi_\alpha$ and $v$, respectively. These two distribution functionals are related by
\begin{eqnarray}
\label{equalityoftwofunctionals}
P[\psi,u,t]=\sum_{\alpha=1}^M\int dv \int_0^\infty d\tau P[\psi,u,t;\psi_\alpha,v,\tau],
\end{eqnarray}
where the integration $\int dv$ is over the whole space of the random variable $u(t)$.

Let $h$ be a small time interval. We can write a probability formula about the detailed probability distribution functional as follows:
\begin{eqnarray}
\label{probrelationbetweentwotimes}
P[\psi,u,t+h;\psi_\alpha,v,\tau+h]&=&\int D\phi D\phi^* dw P[\phi,w,t;\psi_\alpha,v,\tau]\left(1-\Gamma[\phi]h\right)\delta[\psi-\phi+ iG[\phi]h]\nonumber \\
&&\times \delta(u-w-V[\phi]h )
 +{\it o}(h).
\end{eqnarray}
The reason for this is that if the quantum system starts with the target state $\psi_\alpha$ with a value $v$ of the integral~(\ref{integralfunctional}), and successively evolves $\tau+h$ at time $t+h$, the system must also evolve $\tau$ at time $t$, and no jumps occur during the interval $h$. We note that both the Dirac functional $\delta[$ $]$ and function $\delta($ $)$ therein are the consequences of the deterministic Eq.~(\ref{nonlinearSchrodingerequation}). We expand both sides of Eq.~(\ref{probrelationbetweentwotimes}) in terms of $h$ until its first order:
\begin{eqnarray}
&&P[\psi,u,t;\psi_\alpha,v,\tau]+h\partial_t P[\psi,u,t;\psi_\alpha,v,\tau]+h\partial_\tau  P[\psi,u,t;\psi_\alpha,v,\tau]\nonumber \\
&=&\int D\phi  D\phi^*dw  P[\phi,w,t;\psi_\alpha,v,\tau] \delta[\psi-\phi]\delta(u-w)-h\int D\phi  D\phi^*dw P[\phi,w,t;\psi_\alpha,v,\tau]\Gamma[\phi]\delta[\psi-\phi]\delta(u-w)+\nonumber \\
&&ih \int D\phi  D\phi^*dw  P[\phi,w,t;\psi_\alpha,v,\tau]\left\{\int dz\left(\frac{\delta}{\delta \psi(z)}\delta[\psi-\phi]\right) G[\phi](z)\right.-\nonumber \\
&&\left.\int dz \left(\frac{\delta}{\delta \psi^*(z)}\delta[\psi-\phi]\right) G^*[\phi](z)\right\}\delta(u-w)-\nonumber\\
&&h \int D\phi  D\phi^* dw P[\phi,w,t;\psi_\alpha,v,\tau]\delta[\psi-\phi]\left(\frac{\partial}{\partial u}\delta(u-w)\right)V[\phi].
\end{eqnarray}
Letting $h$ tend to zero and using the properties of the Dirac delta function and functional, we obtain
\begin{eqnarray}
\label{quantumsubLiouvillemasterequation}
&&\partial_t P[\psi,u,t;\psi_\alpha,v,\tau]+\partial_\tau  P[\psi,u,t;\psi_\alpha,v,\tau]\nonumber \\
&=&i\int dz \frac{\delta}{\delta \psi(z)} P[\psi,u ,t;\psi_\alpha,v,\tau]   G[\psi](z)-i \int dz \frac{\delta}{\delta \psi^*(z)}  P[\psi ,u,t;\psi_\alpha,v,\tau]   G[\psi]^*(z) -\nonumber \\
&& \Gamma[\psi] P[\psi,u,t;\psi_\alpha,v,\tau]-V[\psi]\frac{\partial}{\partial u} P[\psi,u,t;\psi_\alpha,v,\tau].
\end{eqnarray}

On the other hand, $P[\psi,u,t;\psi_\alpha,v,\tau=0]$ is also equal to the probability distribution in which other wave functions jump to $\psi_\alpha$ at time $t$, and these functions may start with different target wave functions with various successive times of evolution. Hence, we can reformulate it as
\begin{eqnarray}
\label{initialconditionprocess}
P[\psi,u,t;\psi_\alpha,v,0 ]&=& \sum_{\beta=1}^M \int D\phi D\phi^*  dwdv'\int_0^\infty d\tau P[\phi,w,t;\psi_\beta,v',\tau ] k_\alpha[\phi] \delta(w-u) \delta(u-v)\times \nonumber \\
&&\delta \left [\frac{A_\alpha\phi}{\parallel A_\alpha\phi\parallel}-\psi_\alpha\right]
\delta [\psi-\psi_\alpha].
\end{eqnarray}
Eq.~(\ref{initialconditionprocess}) appears to be somewhat lengthy. Nevertheless, the mean of probability is clear.
According to Eq.~(\ref{equalityoftwofunctionals}), by integrating and summing Eq.~(\ref{quantumsubLiouvillemasterequation}) over $\tau$ and $\alpha$, respectively, we obtain a differential equation for the probability distribution functional $P[\psi,u,t]$ as follows:
\begin{eqnarray}
\label{differentialequforPpsiut}
\partial_t P[\psi,u,t]&=&i\int dz\left( \frac{\delta}{\delta \psi(z)}G[\psi](z)-\frac{\delta}{\delta \psi^*(z)}G[\psi]^*(z) \right) P[\psi,u,t]\nonumber -V[\psi]\frac{\partial}{\partial u}P[\psi,u,t]\nonumber \\
&&-\Gamma[\psi]P[\psi,u,t] + \int D\phi D\phi^*  P[\phi,u,t]W[\phi|\psi].
\end{eqnarray}

We are close to the Feynman-Kac formula. Defining Laplace transform of $P[\psi,u,t]$ with respect to the variable $u$,
\begin{eqnarray}
\hat K[\psi,t]= \int du e^{-\lambda u}P[\psi,u,t],
\end{eqnarray}
and using Eq.~(\ref{differentialequforPpsiut}), we derive an equation for the new functional:
\begin{eqnarray}
\label{FeynmanKacformula}
\partial_t \hat K[\psi,t]&=&i\int dz\left( \frac{\delta}{\delta \psi(z)}G[\psi](z)-\frac{\delta}{\delta \psi^*(z)}G[\psi]^*(z) \right )\hat K[\psi,t]\nonumber +\nonumber \\
&&\int D\phi D\phi^* ( \hat K[\phi,t]W[\phi|\psi]-\hat K[\psi,t]W[\psi|\phi] ) -\lambda V[\psi]\hat K[\psi,t].
\end{eqnarray}
Because the Laplace transform $\Phi(\lambda)$ of the probability distribution function $p(u,t)$ for the integral~(\ref{integralfunctional}) is simply related to $\hat K[\psi,t]$ as
\begin{eqnarray}
\label{trajectoryPhi}
\Phi(\lambda)=\int du e^{-\lambda u}p(u,t)=\int D\psi D\psi^*  \hat K[\psi,t],
\end{eqnarray}
if the latter is solved by Eq.~(\ref{FeynmanKacformula}), we can calculate $p(u,t)$ by performing an inverse Laplace transform of $\Phi(\lambda)$. Finally, because $p(u,t)$ has an expression of an ensemble average of the trajectories, Eq.~(\ref{trajectoryPhi}) also leads to
\begin{eqnarray}
\label{trajectoryexpressionFK}
\hat K[\psi,t]=\left \langle \delta[\psi-\psi(t)] e^{-\lambda \int_0^t ds V[\psi(s)]}\right \rangle.
\end{eqnarray}
We name the collection of Eqs.~(\ref{FeynmanKacformula}) and~(\ref{trajectoryexpressionFK}) as the Feynman-Kac formula of the PDP in the Hilbert space.  This is a quantum counterpart of the classical version in the space of classical states~\cite{Davis1984}.

\section{Tilted Liouville-master equation }
\label{section5}
In quantum stochastic thermodynamics, the statistics of heat are of great interest and play a central role~\cite{Esposito2009}. According to the interpretation of quantum thermodynamics, the random jumps of wave functions along quantum trajectories indicate that discrete amounts of heat (the energy quanta) are released to or absorbed from the environment surrounding the quantum systems~\cite{Kurchan2000,Breuer2003, Derezinski2008,Campisi2011,Horowitz2012,Hekking2013,Liu2016a,Liu2018,Liu2020}. We write the heat in a general form:
\begin{eqnarray}
\label{heatdefinition}
Q[\Psi_T]&=&\sum_{i=1}^N \omega_{\alpha_i}=t\sum_{\alpha=1}^M \omega_{\alpha} \left(\frac{1}{t}\sum_{i=1}^N \delta_{\alpha_i,\alpha}\right),
\end{eqnarray}
where $\omega_{\alpha_i}$ are some energy constants that are determined by the target wave functions at time points $t_i$ of the jumps, {\it e.g.}, the Bohr frequencies~\cite{Breuer2000,Liu2016a}. The whole term in the circle brackets of Eq.~(\ref{heatdefinition}) is similar  to Eq.~(\ref{empiricalflux}). It is indeed an integration of the latter, that is,
$\int D\phi D\phi^* F_{\phi,\alpha}[\Psi_t]$.

Now, we can construct a differential equation for the heat. Analogous to previous cases, we focus on its Laplace transform (or the moment generating function) rather than distribution: $\Xi(\lambda)=\left\langle \exp(-\lambda Q)\right\rangle $, and the right-hand side of the equation is
\begin{eqnarray}
\label{Laplacetransformheat}
\sum_{\alpha_0,\Psi_t} P[\psi_{\alpha_0}] \exp\left (-t \int D\phi D\phi^* \Gamma[\phi]P_{\phi}[\Psi_t]+t \sum_{\alpha=1}^M\int D\phi D\phi^* \ln \left( e^{-\lambda\omega_\alpha}k_\alpha[\phi]\right) F_{\phi,\alpha}[\Psi_t]  \right).
\end{eqnarray}
Eq.~(\ref{Laplacetransformheat}) inspires us to define an auxiliary open quantum system as follows: its nonlinear Schr$\ddot{o}$dinger equation is the same as Eq.~(\ref{nonlinearSchrodingerequation}), but the rates of the jumps of wave functions are modified to
\begin{eqnarray}
\label{auxiliaryrate}
k'_\alpha[\phi] =e^{-\lambda\omega_\alpha}k_\alpha[\phi].
\end{eqnarray}
In the remainder of this paper, we always denote quantities in the auxiliary open quantum system with a prime symbol unless stated otherwise. With these notations, we can rewrite the Laplace transform of the heat as
\begin{eqnarray}
\label{heatmomentgeneratingfunction}
\Xi(\lambda)&=&\sum_{\alpha_0,\Psi_t} P[\psi_{\alpha_0}]\mathbf{P'}_{0}[\Psi_t]\frac{\mathbf{P}_{0}[\Psi_t]}
{\mathbf{P'}_{0}[\Psi_t]}e^{-\lambda Q[\Psi_t]} \nonumber \\
&=&\int D\phi D\phi^*\left\langle \delta[\phi-\psi(t)] e^{-\int_0^t ds (\Gamma[\psi(s)]-\Gamma'[\psi(s)])}\right\rangle',
\end{eqnarray}
We note that the probability distribution functional $\mathbf{P'}_{0}[\Psi_t]$ of monitoring the trajectory $\Psi_t$ is the same as Eq.~(\ref{probdensityfunctionaloftrajectory}) except that the rates therein are replaced by Eq~(\ref{auxiliaryrate}). So is the total rate $\Gamma'[\psi]$; see Eq.~(\ref{totalrate2}). Defining the integrand of Eq.~(\ref{heatmomentgeneratingfunction}) as $\hat P [\psi,t]$ and using the Feynman-Kac formula~(\ref{FeynmanKacformula}) and~(\ref{trajectoryexpressionFK}), we immediately obtain
\begin{eqnarray}
\label{quantumtiltinggeneratorlevel1}
\partial_t \hat P [\psi,t]&=&i\int dx\left ( \frac{\delta}{\delta \psi(x)}G[\psi](x)-\frac{\delta}{\delta \psi^*(x)}G[\psi]^*(x) \right )\hat P [\psi,t] \nonumber +\nonumber \\
&&\int D\phi D\phi^* \left ( \hat P [\phi,t]W'[\phi|\psi]-\hat P [\psi,t]W[\psi|\phi] \right ),
\end{eqnarray}
where
\begin{eqnarray}
\label{rateauxiliarysystem}
W'[\phi|\psi]=\sum_{\alpha=1}^M e^{-\lambda \omega_\alpha} k_\alpha[\phi]\delta\left[\frac{A_\alpha \phi }{\parallel A_\alpha \phi\parallel }-\psi\right].
\end{eqnarray}
We name Eq.~(\ref{quantumtiltinggeneratorlevel1}) the tilted Liouville-master equation, which is the central result of this paper. If $\lambda$ is zero, the equation yields the Liouville-master equation~(\ref{quantumLiouvillemasterequation}). Hence, we have definitely answered the first two questions in Fig.~(\ref{fig1}).

\subsection{Tilted quantum master equation}
Here, we demonstrate that Eq.~(\ref{quantumtiltinggeneratorlevel1}) can yield the tilted quantum master equation~\cite{Esposito2009}, {\it  i.e.}, the third question of this paper.  The procedure is very similar to that of the Liouville-master equation and quantum master equation~\cite{Breuer1995a,Breuer1995,Breuer1997,Breuer1997a}. Let us consider an operator $\hat\rho(t)$ that is defined as
\begin{eqnarray}
\label{heatproductionrhozz}
\hat{\rho}(z,z',t)=\int D\psi D\psi^* \hat P [\psi,t]\psi(z)\psi^*(z')
\end{eqnarray}
in the position representation. Taking the time derivatives of its both sides and substituting Eq.~(\ref{quantumtiltinggeneratorlevel1}), we have
\begin{eqnarray}
\label{partialhatrho}
\partial_t\hat{\rho}(z,z',t)
&=&i \int D\psi D\psi^*\int dx \psi(z)\psi^*(z')\left( \frac{\delta}{\delta \psi(x)}G[\psi](x)-\frac{\delta}{\delta \psi^*(x)}G[\psi]^*(x) \right) \hat P[\psi,t]\nonumber +\nonumber \\
&& \int D\psi D\psi^* \psi(z)\psi^*(z')\int D\phi D\phi^*\left( \hat P[\phi,t] W'[\phi|\psi] -\hat P[\psi,t]W[\psi|\phi] \right).
\end{eqnarray}

There are four terms on the right-hand side of Eq.~(\ref{partialhatrho}). The first term is equal to
\begin{eqnarray}
\label{eq1}
&&-i \int D\psi D\psi^*\int dx G[\psi](x)   \hat P[\psi,t]\left(\frac{\delta}{\delta \psi(x)}\psi(z)\psi^*(z')\right)\nonumber \\
&=&-i  \int D\psi D\psi^*\hat P[\psi,t]\psi^*(z') G[\psi](z)   \nonumber \\
&=&-i  \int dy\int D\psi D\psi^*  \hat P[\psi,t]\psi^*(z')\langle z|\left(H-\frac{i}{2}\sum_{\alpha=1}^M r_\alpha A_\alpha^\dag A_\alpha + \frac{i}{2}\sum_{\alpha=1}^M r_\alpha \parallel A_\alpha\psi\parallel ^2\right)|y\rangle \psi(y)\nonumber \\
&=&-i \int dy  \langle z|H|y\rangle \hat\rho(y,z',t) - \frac{1}{2}\sum_{\alpha=1}^M r_\alpha\int dy  \langle z|A_\alpha^\dag A_\alpha|y\rangle \hat\rho(y,z',t) + \nonumber \\
&&\frac{1}{2}\sum_{\alpha=1}^M r_\alpha \int D\psi D\psi^*  \hat P[\psi,t]\psi(z)\psi^*(z') \parallel A_\alpha\psi\parallel^2.
\end{eqnarray}
The first equation of Eq.~(\ref{eq1}) uses the functional integration by parts. Because of the properties $\delta\psi(z)/\delta\psi(x)=\delta(z-x)$ and $\delta\psi^*(z)/\delta\psi(x)=0$, we arrive at the second equation. The third equation therein is a consequence of inserting Eq.~(\ref{Goperator}). Performing similar calculations, the second term of Eq.~(\ref{partialhatrho}) is
\begin{eqnarray}
&&i \int dy  \hat\rho(z,y,t) \langle y|H|z'\rangle - \frac{1}{2}\sum_{\alpha=1}^M r_\alpha\int dy  \hat \rho(z,y,t)  \langle y|A_\alpha^\dag A_\alpha|z'\rangle   + \nonumber \\
&&\frac{1}{2}\sum_{\alpha=\alpha}^M r_\alpha \int D\psi D\psi^*  \hat P[\psi,t]\psi(z)\psi^*(z') \parallel A_\alpha\psi\parallel^2.
\end{eqnarray}
By inserting Eq.~(\ref{rateauxiliarysystem}), we can rewrite the third term  as
\begin{eqnarray}
&&\sum_{\alpha=1}^M r_\alpha e^{-\lambda \omega_\alpha} \int D\psi D\psi^* \psi(z)\psi^*(z')\int D\phi D\phi^*  \hat P[\phi,t] \parallel A_\alpha \phi \parallel ^2\delta\left[\frac{A_\alpha \phi}{\parallel A_\alpha \phi \parallel} -\psi\right] \nonumber\\
&=&\sum_{\alpha=1}^M r_\alpha e^{-\lambda \omega_\alpha} \int D\phi D\phi^*   \langle z|A_\alpha|\phi\rangle  \hat P[\phi,t]  \langle \phi |A^\dag_\alpha|z'\rangle \nonumber \\
&=&\sum_{\alpha=1}^Mr_\alpha e^{-\lambda \omega_\alpha} \int dydy' \langle z|A_\alpha|y\rangle \hat\rho(y,y',t) \langle y' |A^\dag_\alpha|z'\rangle.
\end{eqnarray}
The fourth term is relatively simple:
\begin{eqnarray}
\label{eq4}
-\sum_{\alpha=1}^M r_\alpha \int D\psi D\psi^*  \hat P[\psi,t]\psi(z)\psi^*(z') \parallel A_\alpha \psi  \parallel^2.
\end{eqnarray}
Eqs.~(\ref{eq1})-(\ref{eq4}) appears tedious. However, when we substitute them into the right-hand side of Eq.~(\ref{partialhatrho}), we find that this differential equation leads to a concise form for the operator $\hat\rho(t)$:
\begin{eqnarray}
\label{tiltedquantumasterequation}
\partial_t \hat \rho(t)=-i[H,\hat\rho(t)]+\sum_{\alpha=1}^M r_\alpha\left ( e^{-\lambda\omega_\alpha}A_\alpha\hat \rho(t)A^\dag_\alpha -\frac{1}{2}\left\{ A^\dag_\alpha A_\alpha,\hat\rho(t) \right \} \right ).
\end{eqnarray}
Accordingly, an alternative expression for the Laplace transform of the heat is obtained: $\Xi(\lambda)={\rm Tr}[\hat \rho(t)]$. Eq.~(\ref{tiltedquantumasterequation}) is nothing but TQME~\cite{Esposito2009}. Analogous to the relationship between Eqs.~(\ref{quantummasterequation}) and (\ref{quantumLiouvillemasterequation}), we name Eq.~(\ref{quantumtiltinggeneratorlevel1}) the unraveling of Eq.~(\ref{tiltedquantumasterequation}) in Hilbert space. Because of the absence of the abstract functional derivatives, the latter is more advantageous for the calculation and analysis of the statistics of random heat~(\ref{heatdefinition}).

\section{Conclusion}
\label{section6}
We have obtained the tilted Liouville-master equation in the Hilbert space and this equation can yield the tilted quantum master equation of open quantum systems.

\begin{acknowledgments}
We appreciate Prof. Hong Qian and Dr. Carollo for their inspiring discussions. This work was supported by the National Science Foundation of China under Grant No. 11174025 and No. 11575016.
\end{acknowledgments}

\end{document}